# Luminescent Ge-related centre in high-pressure synthesized diamond.

E.A. Ekimov[1], S.G. Lyapin[1], K.N. Boldyrev[2], M.V. Kondrin[1], R. Khmelnitskiy[3], V.A. Gavva[4], T.V. Kotereva[4], M.N. Popova[2]

[1]Institute for High Pressure Physics, Russian Academy of Sciences, 142190 Troitsk, Moscow, Russia

[2]Institute of Spectroscopy, Russian Academy of Sciences, 142190 Troitsk, Moscow, Russia

[3]Lebedev Physics Institute, Russian Academy of Sciences, 117924 Moscow, Russia

[4]Institute of Chemistry of High-Purity Substances, Russian Academy of Sciences, 603950 Nizhny Novgorod, Russia

**Colour centres in diamond promise attractive applications in biology, magnetometry, nanoscopy, and in different modern quantum information processing and communication (QIPC) technologies[1-3]. On the basis of nitrogen-vacancy centres, qubits manipulated by optical and microwave radiation and entanglement of two distant qubits were demonstrated[4,5] and optical quantum memories were proposed[6,7]. Colour centres in diamond can serve as single-photon emitters[8-11]. For this and other QIPC applications, narrow optical transitions and reproducible properties of colour centres can be achieved on the base of diamond with high structural perfection. Here, we report on the high-pressure synthesis of novel nano- and microcrystalline high-quality diamonds with luminescent Ge-related centres. Observed germanium and carbon isotope shifts in the fine structure of luminescence at 2 eV (602 nm) allow us to unambiguously associate the nature of the centre with the germanium impurity in diamond. High-intensity and narrow-line emission at 2 eV makes the Ge-related centre in high-pressure synthesized nanodiamonds a very prospective candidate for single-photon emitters and other QIPC applications.**

During the last decade, colour centres in diamond have become an object of very intense study in the context of their possible use as luminescent markers in biology [1,2], for measuring weak magnetic fields with high spatial resolution[1,3] and, the most important, as three-level Λ systems[4-7] and single-photon emitters[8-11] for quantum information processing and communication (QIPC) technologies. In particular, nitrogen-vacancy[8] (N-V), silicon-vacancy[9] (Si-V), Cr-related[10] and Ni-related[11] (NE8) centres in diamond were shown to be capable of single-photon emitting. The search and study of new photon-luminescent centres in diamond is of great importance not only for expanding the spectral range of single-photon emitters, but also for understanding the relationship between the structure and energy level scheme and luminescent properties of colour centres in diamond, which is essential for designing an efficient platform for QIPC. In the theoretical paper by Goss[12], it was predicted that germanium, an element with larger than for silicon atomic radius, can form a Ge-V colour centre in diamond with structure similar to that of the Si-V centre and emit photons in the same spectral range. Recently, luminescence at a wavelength of about 602 nm, different from the near-infrared spectral region of the Si-V centre, was reported for diamond doped with germanium by ion implantation and in the process of chemical vapour deposition (CVD)[13]. It was demonstrated that the "602 nm" centre works as a single-photon source, whereas the nature of the centre was not actually established in the presence of other impurities. Luminescence at "602 nm" of CVD diamond remained broad even at 5 K indicating imperfect structure of the center[13,14].

High-pressure, high-temperature (HPHT) synthesis is historically known for producing diamond crystals of a very high optical quality. For instance, fine structure in the luminescence line of the Si-V centre at 737 nm was first investigated on Si-doped diamond crystals grown by the temperature-gradient method in metallic melts[15,16]. In practical terms, the size of diamond crystals with single-photon emitters is desirable to be less than the emission wavelength to provide an optimal performance of the emitter[17]. In this regard, HPHT technology was shown to be very effective for the massive synthesis of pure[18,19] and doped[20,21] nanodiamonds from organic compounds. As for germanium impurity, nothing is known about its presence in diamond grown at high pressures, even from a germanium melt[22]. In this context, HPHT synthesis of Ge-doped diamonds becomes very important not only for elucidating the nature of the "602 nm" luminescent centre, but also for producing a platform of appropriate size for new single-photon emitters. It is also worth mentioning that the $^{73}$Ge isotope possesses a nonzero nuclear spin ($I = 9/2$) and, thus, a hyperfine structure of energy levels, which could be used to build a three-level $\Lambda$ system for some QIPC applications.

Here, we describe HPHT synthesis of small high-quality diamond crystals with Ge-related colour centres in the C-H-Ge growth system. Observation of the four-line fine structure in luminescence at 602 nm at temperatures below 80 K manifests a high quality of HPHT diamonds. We show that there are two ground-state energy levels with the separation of 0.7 meV and two excited-state levels separated by 4.6 meV in the electronic structure of the centre. Similar electronic structures of the Ge- and Si-related colour centres, as well as our DFT calculations (performed on hexagonal supercell of 84 atoms with the [0001] direction parallel to the <111> direction of diamond lattice) strongly support a similarity of these two centres in the crystal structure.

HPHT diamond synthesis from hydrocarbons became the basis for producing small diamond crystals with a high structural perfection. Naphthalene, $C_{10}H_8$, mixed with germanium was used for the synthesis of germanium-doped diamonds (Figure 1). Similar to the synthesis of diamond from pure naphthalene[23], diamond formation in the Ge-C-H growth system at 8-9 GPa takes place at temperatures above 1600 K. At these temperatures germanium does not act as a catalyst for the diamond synthesis[22], which makes it possible to control doping of diamond by changing concentration of germanium in the growth system. All crystals obtained were transparent; neither non-diamond phases nor metallic impurities were detected in acid-treated powder samples by x-ray phase and EDX analysis (Figure S1, Supplementary Information).

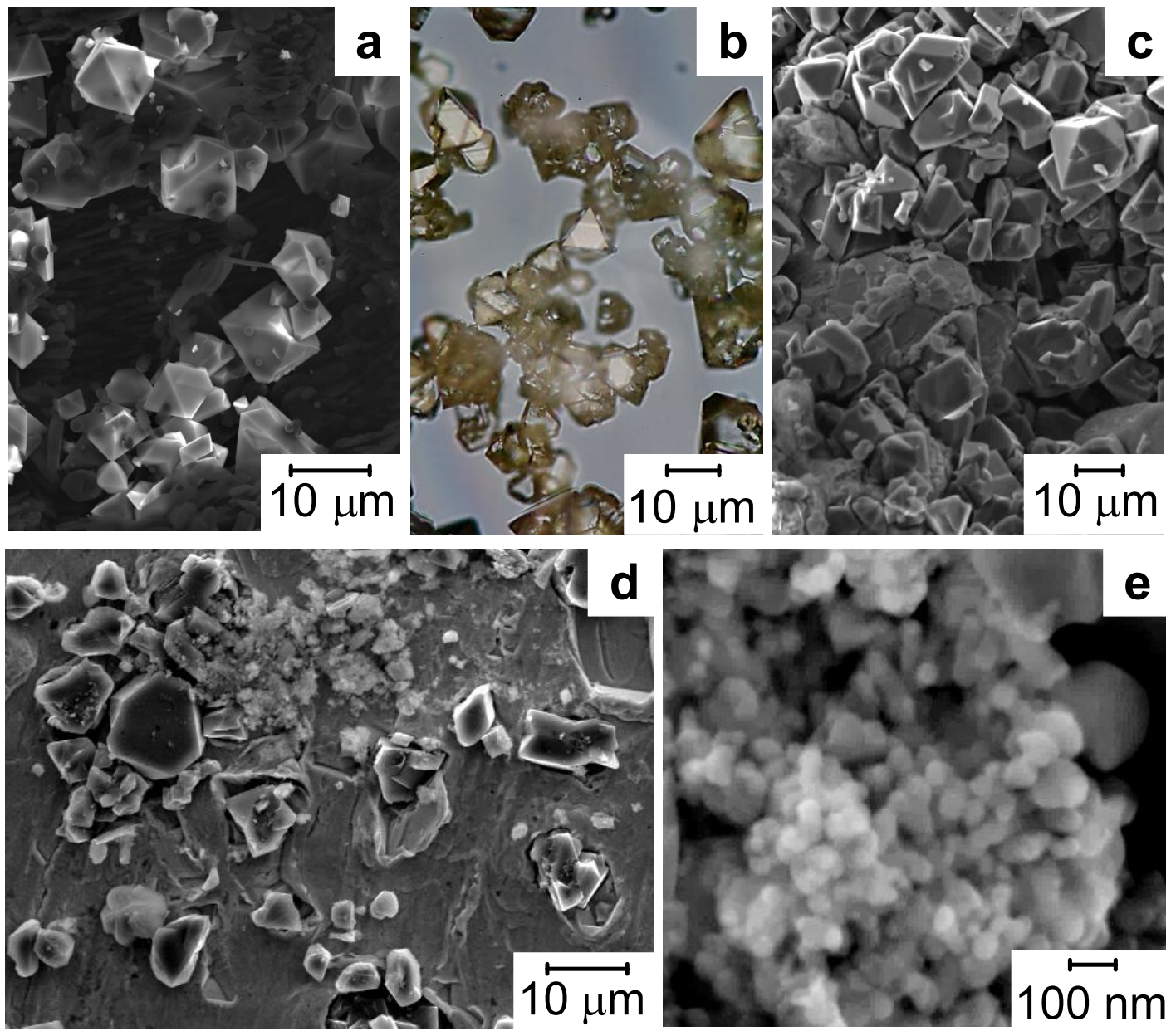

**Figure 1 | Diamonds synthesized at high pressures in the C-H-Ge system. a**,**b**, Microcrystals of up to 10-15 μm in size with perfect shape were obtained at pressure 7-8 GPa. **c**, **d**, **e**, At higher pressure, 8-9 GPa, the occurrence of diamond nanocrystals of 30 to 100 nm in size was revealed among microcrystalline diamonds. Microdiamonds as-synthesized (**a**) and after acid treatment in **b**; micro- and nano-sized as-synthesized diamonds (**c**) and after acid treatment on the indium plate in **d**; nano-sized diamonds (enlarged part taken from the previous image) in **e**. (**a, c, d, e**) – scanning electron microscopy images, (**b**) - the optical microscope image.

Figure 2a shows typical photoluminescence (PL) spectra collected on diamonds doped with germanium of natural isotope composition (nGe). At 300K, a zero-phonon line (ZPL) was found at 602.5 nm with FWHM 4.5 nm (~16 meV). The ZPL, typically 2-10 times stronger in intensity than the diamond Raman line, becomes more intense with increase of the germanium concentration in the growth medium. At 80K, the ZPL splits into two lines at 602.1 and 601.1 nm with FWHM 0.25 nm (~0.9 meV). Further cooling to 10 K results in a significant narrowing of the lines to 0.1nm (~0.38meV). Inspecting vibronic sideband structure of the ZPL at 10 K, we have found only one sharp line at 615 nm (45.2 meV with respect to the ZPL) (Figure 2b). Depending on the synthesis conditions, $(Si\text{-}V)^-$ line, $(N\text{-}V)^0$ and $(N\text{-}V)^-$ lines could be detected in the spectra, see inset of Figure 2a and Figure 2b. The presence of silicon in the growth system was caused by contaminating the initial mixture during its preparation in a jasper mortar. In contrast to the Ge-doped diamond films[13,14], the presence of silicon impurity in our Ge-doped diamonds was easily eliminated by using a plexiglass mortar instead of a jasper one (Figure S2, Supplementary Information). The presence of $(N\text{-}V)^0$ and $(N\text{-}V)^-$ peaks in the luminescence spectra of diamonds was revealed at cooling down to 80 K only. Interestingly that boron, if added to the growth system, lowers nitrogen concentration in diamond to a level undetectable in luminescence spectra, as well as decreases FWHM of the luminescent line at 602 nm. Thus, simple manipulations in chemical composition of starting mixtures make it possible to govern the luminescence spectrum of HPHT diamonds.

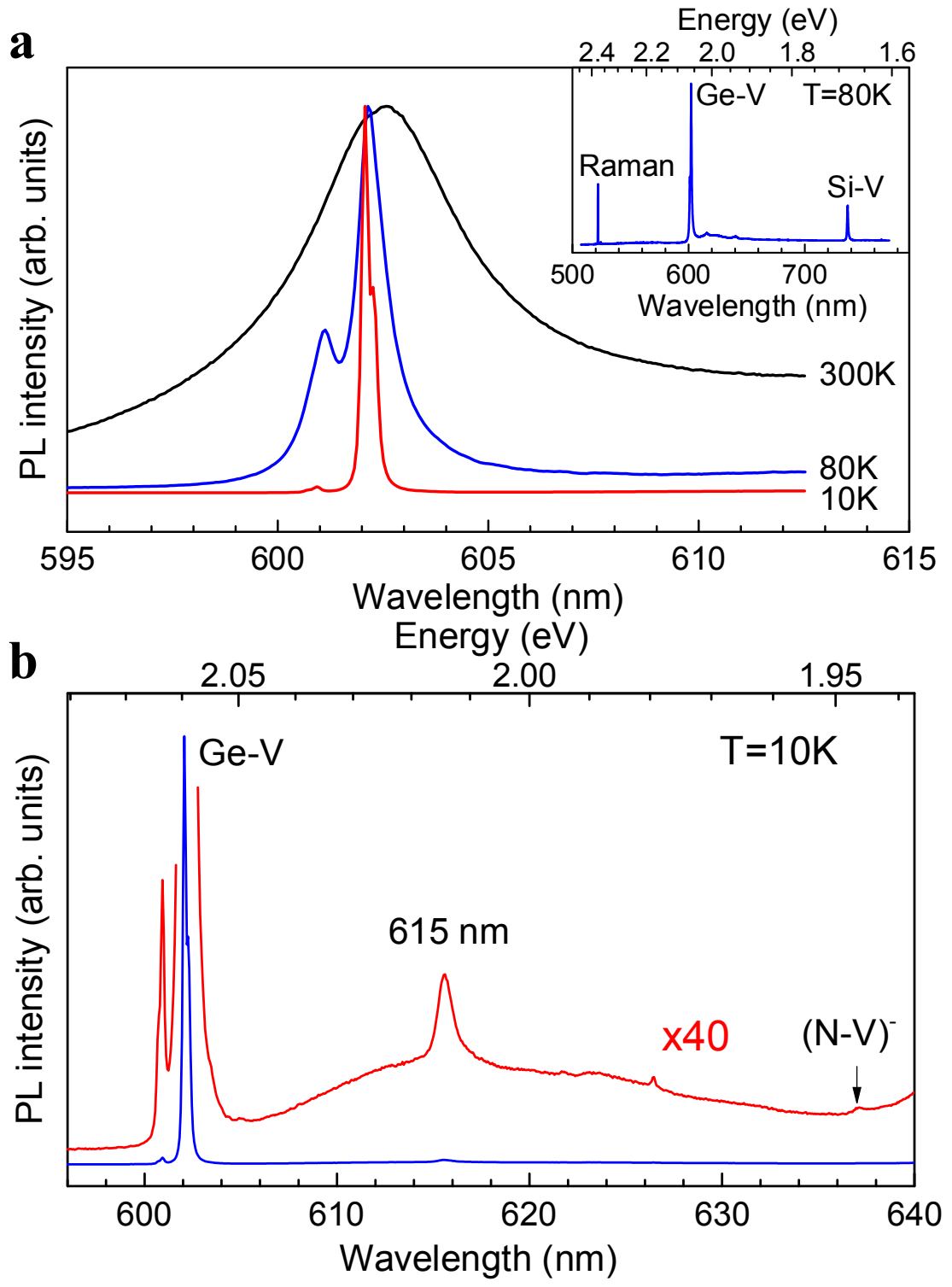

**Figure 2 | Photoluminescence spectra of Ge-doped diamond. a,** Splitting of the "602 nm" line when cooled from 300K to 80K and 10K. **Inset** shows a PL spectrum at 80K with all three dominating features: ZPL at 602 nm, the Raman peak from diamond and the Si-V line at 737 nm. **b,** Sideband structure near the ZPL revealing 615 nm localized vibrational mode besides the $(N\text{-}V)^-$ peak at 637 nm. The sideband intensity is multiplied by a factor 40.

High-resolution spectra recorded at 10 K reveal the four-line fine structure of ZPL (Figure 3a), which can be interpreted in terms of optically allowed transitions between the doublet ground and excited states (Figure 3c) in the same manner as for the fine structure of ZPL in the case of the Si-V center[16]. In total, more than twenty micro-sized (3-9 µm) and aggregated nano-sized (30-100 nm) diamonds were studied at 80 K. No significant variation in the position of ZPL was detected, while FWHM of the ZPL became smaller at low germanium concentration.

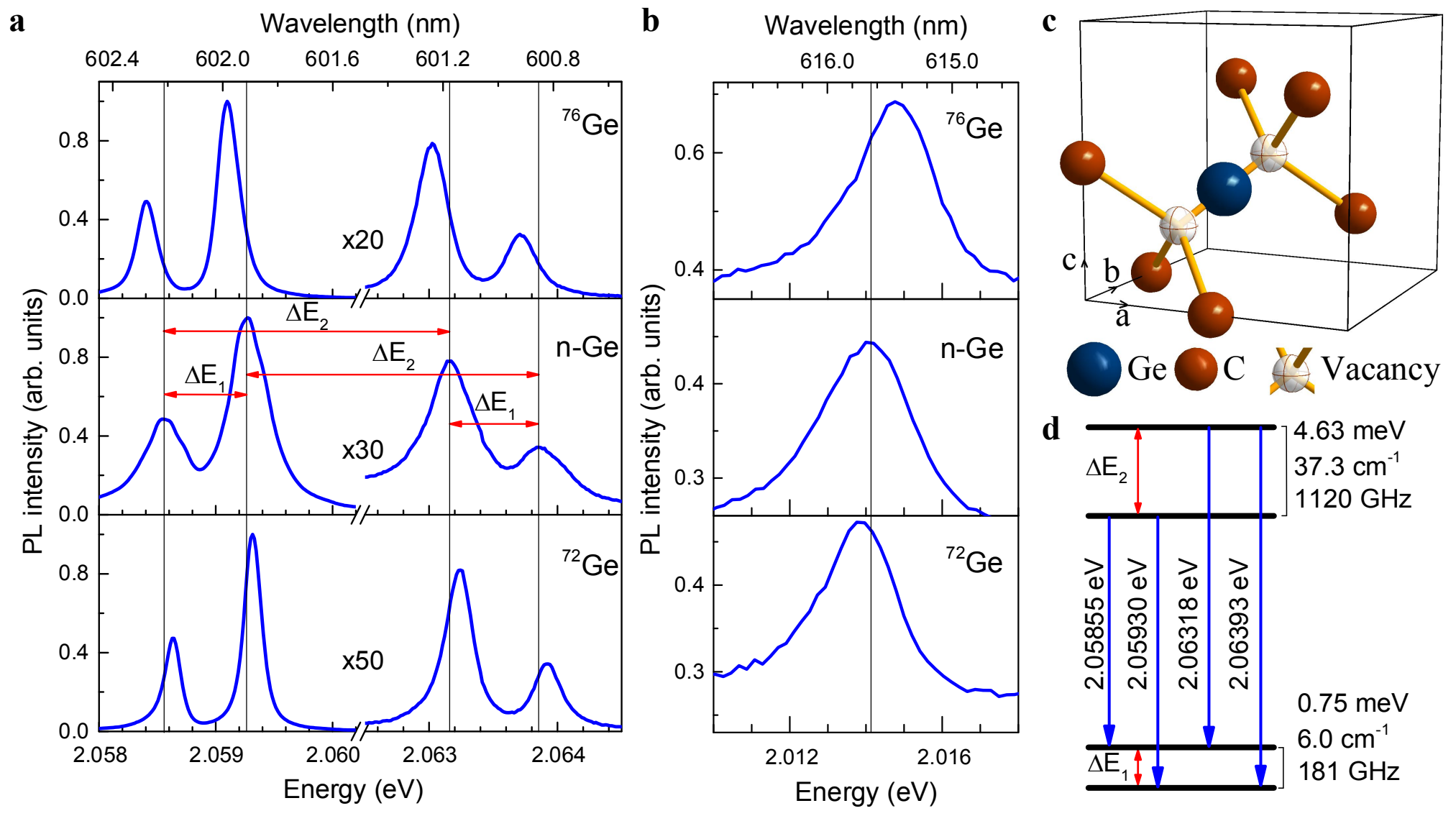

**Figure 3 | Isotopically varying spectral features of Ge centre in diamond. a,** High-resolution spectra of the 602 nm Ge-related peak reveal 2 doublets, which shift to lower energy as the Ge atom increases in mass. Note that replacing natural Ge by isotopic-pure $^{72}$Ge results in an extraordinary narrowing of ZPL down to 0.05 nm (~0.16 meV). **b**, The sharp phonon peak at 615 nm moves to higher energy as the Ge mass increases. The actual energy of the phonon mode is given by the separation between this peak and the zero phonon line, and therefore the phonon energy decreases for nGe and $^{76}$Ge. **c**, The split-vacancy configuration of the Ge-V centre with the Ge atom (blue) in between the unoccupied lattice sites (white) and the nearest-neighbour carbon atoms (red). **d**, Energy-level scheme of the Ge-V centre. The split ground and excited states give rise to four optical transitions presented at panel **a**.

To prove that ZPL at 602 nm and the localized vibrational mode (LVM) at 615 nm originate from the Ge centre we have grown samples *i)* replacing natural Ge by the $^{72}$Ge or $^{76}$Ge isotopes and *ii)* replacing natural carbon by the $^{13}$C isotope. In the case *i),* we observed that all four peaks of the ZPL shift to longer wavelengths as the Ge atom increases in mass (Figure 3a), which unambiguously associates the nature of the centre with the impurity of germanium in diamond. Meanwhile, the distance between the ZPL and the LVM diminishes, in agreement with the phonon energy decrease for a heavier atom. The measured phonon energies for this feature are $E_{LVM(72)}$=45.43±0.01 meV for $^{72}$Ge, $E_{LVM(nGe)}$=45.24±0.01 meV for nGe, and $E_{LVM(76)}$=44.31±0.01 meV for $^{76}$Ge. We calculated the ratios of the LVM energies and compared them with the ratio of the isotope masses, following the direction of Ref. [20] for the Si-V centre.

$$\frac{E_{LVM(nGe)}}{E_{LVM(76)}} = 1.0210 \pm 0.0003 \sim 1.0207 = \sqrt{\frac{m_{76}^*}{m_{nGe}^*}}, \text{ and}$$

$$\frac{E_{LVM(72)}}{E_{LVM(76)}} = 1.0254 \pm 0.0003 \sim 1.0255 = \sqrt{\frac{m_{76}^*}{m_{72}}}.$$

Here, $m_{76}^*$ and $m_{nGe}^*$ are the effective masses of the used germanium isotopes (see About the effective masses in Supplementary Information). The measured ratios are in close agreement with a simple harmonic oscillator model where the phonon frequency $\omega$ is given by $\omega = \sqrt{\frac{k}{m}}$. The spring constant $k$ and an oscillating mass $m$ refer to the germanium atom. The validity of this simple approach indicates that the 615 nm spectral feature arises from oscillations of the Ge atom, carbon atoms of the diamond lattice being not much involved.

This conclusion is further supported by our results on the samples *ii)*. On changing the $^{13}C$ content from 1.1% in natural carbon to 99%, we observed the isotope shift +2.8 meV for the ZPL (Figure 3S, Supplementary Information), while the energy of the LVM does not noticeably change with carbon isotopic substitution. These observations are consistent with a model of a strongly localized state of a heavy Ge atom in the diamond lattice.

Using DFT calculations performed on a hexagonal supercell of 84 atoms with the [0001] direction parallel to the <111> direction of the diamond lattice, we show that the structure of the Ge-V defect is consistent with the split-vacancy model proposed for the Si-V defect[25,26]. Upon relaxation of atomic positions in the supercell, a germanium atom moves halfway in the direction to the nearby vacant cite. The resulting band structure near the top of the valence band in the Γ point of Brillouin zone is schematically shown in Figure 4. In this energy range, two doubly-degenerate defect levels are present. Experimentally observed photoluminescence can be explained by transitions between partially occupied level deep into band gap and the one close to the top of the valence band. For the $(Ge\text{-}V)^{-1}$ centre, the energy-level difference was calculated to be 1.91 eV, whereas the electron excitation energy (calculated using inverse population of energy levels) yields higher energy of 2.01 eV, close to that observed experimentally.

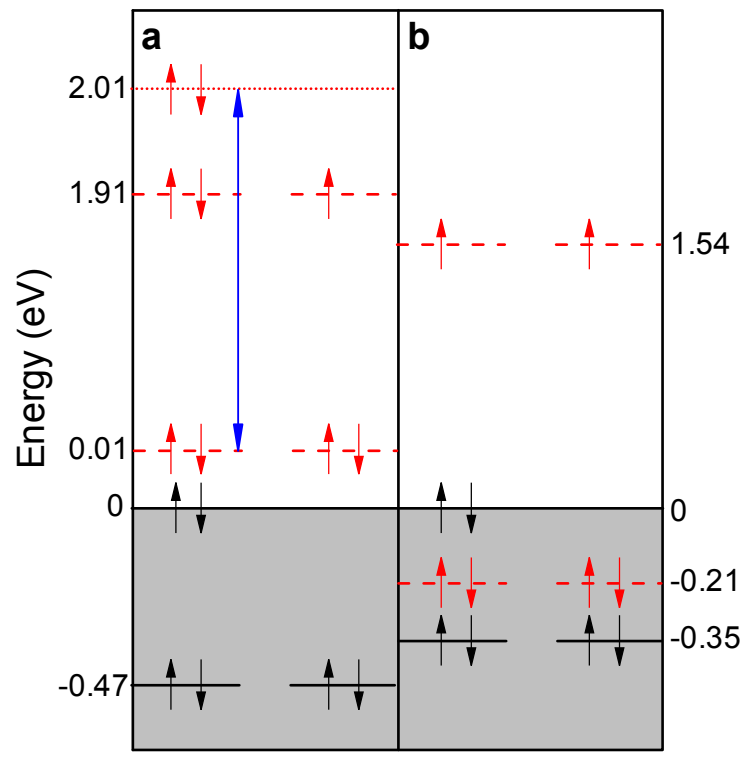

**Figure 4 | Schematic representation of energy levels of the charged impurity centre $(Ge\text{-}V)^{-1}$ (a) and the neutral impurity centre $(Ge\text{-}V)^{0}$ (b) in the vicinity of the valence band maximum.** The grey area represents the valence band. Solid lines stand for a triply-degenerate level at the top of the pure diamond valence band ($T_{2g}$) which is split by the impurity. Dashed lines indicate doubly-degenerate $(Ge\text{-}V)^{x}$ impurity levels (for clarity, we neglect the spin-orbital coupling which is at most of the order of 10 meV). Dotted line at the left panel designates an excited state of the Ge impurity. Arrows show occupation of respective levels by electrons. The double-sided arrow at the left panel demonstrates an electronic transition responsible for the observed photoluminescence.

The energy splitting values due to the spin-orbit coupling were found to be of about 10 meV for the lower energy pair and about 2 meV for the upper one, while Jahn-Teller splitting values were of the order of 1 meV. Since the energy difference below 10 meV is at the brink of possibilities of DFT, values for energy level splitting must be considered as an order of magnitude estimate only

In summary, we demonstrate a great potential of the HPHT technology in preparation of small high-quality diamonds with colour centres, which can be used as a platform for single-photon emitters. By detecting carbon and germanium isotope shifts in the luminescence of diamond at 2 eV and its vibrational sideband, we show that germanium enters into the diamond lattice forming Ge-related colour centre. Similarities in electron structure of Ge- and Si-related colour centres, as well as our DFT calculations strongly support split-vacancy structure of this centre.

***Sample preparation methods***

*Powders of Ge-doped diamond crystals were synthesized with the use of carbon and germanium of natural isotopic composition, 99%* $^{12}C$ *and nGe (36.5%* $^{74}Ge$*, 27.4%* $^{72}Ge$*, 20.5%* $^{70}Ge$*, 7.8%* $^{73}G$*, 7.8%* $^{76}Ge$*), as well as enriched with the isotopes* $^{13}C$ *(99.3 at.%),* $^{72}Ge$ *(99.98%) or* $^{76}Ge$ *(88%* $^{76}Ge$*, 12%* $^{74}Ge$*). Ge-doped diamonds with different germanium isotopic composition were produced from powder mixtures of naphthalene (*$C_{10}H_8$*, 99%* $^{12}C$*) and germanium. Mixtures of naphthalene with different Ge percentage of 0.7, 4 and 13 mass % were prepared in a mixer (mortar and pestle, both made of jasper or plexiglass) for about 5 min, pressed into pellets and placed inside titanium capsules. For diamond synthesis, toroid-type high-pressure chamber was used to generate pressures 7-9 GPa and a desired temperature between 1600 and 1900 K in the reaction volume. Under pressure, total duration of the heat treatment was about 10 c. After the treatment, samples were quenched under pressure to room temperature by switching of the electric power. Then, powdered samples were recovered from the capsules and treated consequently in aqua regia, solution of nitric and sulfuric acids (1:3), and perchloric acid to remove nondiamond phases from the samples. Diamond enriched with isotope* $^{13}C$ *was synthesized from* $^{13}C$ *amorphous carbon. Since synthesis of diamond in the growth system nGe (13 mass %) -* $^{13}C$ *amorphous carbon caused uncontrollable Ge-doping of diamond and trapping of germanium-based inclusions (see Figure S1 in Supplementary Information),* $H_2O$ *(12 mass %) was added to the mixture in order to prevent synthesis of diamond in the double Ge-C system. Unlike base C-H-Ge growth system, crystal structure of diamond synthesized in the presence of water was less perfect.*

***Optical measurements.***

*PL spectra were taken using the 488 nm $Ar^+$ laser line for excitation and a triple-grating spectrometer (Princeton Instruments TriVista 555) with a liquid-nitrogen-cooled CCD detector. For measurements in the temperature range 80-300 K, the samples were placed inside a cryostage (Linkam THMS600) and 50x objective (N.A.=0.50) of a confocal microscope (Olympus BX51) was used for laser focusing and PL signal collection. For measurements at 10K the samples were placed in a He cryostat (Oxford Instruments OptistatSXM) and an achromatic lens was used for focusing and collecting the signal. The laser spot on the sample inside the cryostat was about ~5 µm. Low-resolution spectra covering the entire sideband were measured using a spectrometer in subtractive configuration with a 600-grooves-per-mm grating, which gave resolution ~0.25 meV (~0.1 nm). The ZPL was measured with a 1800- grooves-per-mm grating in additive configuration which gave a ~0.03 meV (~0.01 nm) resolution capable of resolving the fine structure of four lines, separated by ~0.7 meV and ~4.6 meV and with typical FWHM=0.1-0.4 meV.*

***Calculations***

*DFT calculations were carried out with Quantum ESPRESSO software package[27]. Perdew-Burke-Ernzerhof exchange correlation functional was applied with norm-conserving pseudopotentials (of the Trouiller-Martins type) for both carbon and germanium atoms. Energy cut-off was 140 Ry. For integration over the Brillouin zone, unshifted 8×8×8 Monkhorst-Pack grid was used. In the process of calculation, the relaxation of cell dimensions and ion positions was done until the residual force on every atom was less than 0.001 Ry/bohr and the residual stress was less than 50 MPa. For calculation of the spin-orbital splitting, we used already relaxed supercells with substitution of fully relativistic pseudopotential for germanium only. In this case, only self-consistent calculations were carried out. Due to deficiencies of PBE functionals which regularly underestimate band gaps, the band structure was subsequently corrected by using a hybrid functional (HSE06) with standard parameters (0.25 for the mixing ratio and 0.2 $Å^{-1}$ for the screening value).*

**Acknowledgments**

The authors acknowledge financial support from the Russian Academy of Sciences under the Programs for Basic Research, Russian Foundation for Basic Research, Projects 15-02-05603 and 13-02-01091. The authors thank V.G. Ralchenko for helpful discussions.

**Authors' contributions.**

Diamond samples were prepared by E.A.E., photoluminescence measurements were performed by S.G.L. and absorption was studied by K.N.B., germanium isotopes were prepared by V.A.G., T.V. K., and microstructure of samples was investigated by R.K. DFT calculations were carried out by M.V.K. Authors: E.A.E., S.G.L., K.N.B., M.V.K., M.N.P. contributed to analysis of the data, discussion and writing the paper.

**Competing interests statement.**

The authors declare that they have no competing financial interests.

**Correspondence** and requests for materials should be addressed to E.A.E. (ekimov@hppi.troitsk.ru)

# SUPPLEMENTARY INFORMATION

## Luminescent Ge-related centre in high-pressure synthesized diamond.

E.A. Ekimov[1], S.G. Lyapin[1], K.N. Boldyrev[2], M.V. Kondrin[1], R. Khmelnitskiy[3], V.A. Gavva[4], T.V. Kotereva[4], M.N. Popova[2]

[1]Institute for High Pressure Physics, Russian Academy of Sciences, 142190 Troitsk, Moscow, Russia

[2]Institute of Spectroscopy, Russian Academy of Sciences, 142190 Troitsk, Moscow, Russia

[3]Lebedev Physics Institute, Russian Academy of Sciences, 117924 Moscow, Russia

[4]Institute of Chemistry of High-Purity Substances, Russian Academy of Sciences, 603950 Nizhny Novgorod, Russia

### 1. Synthesis

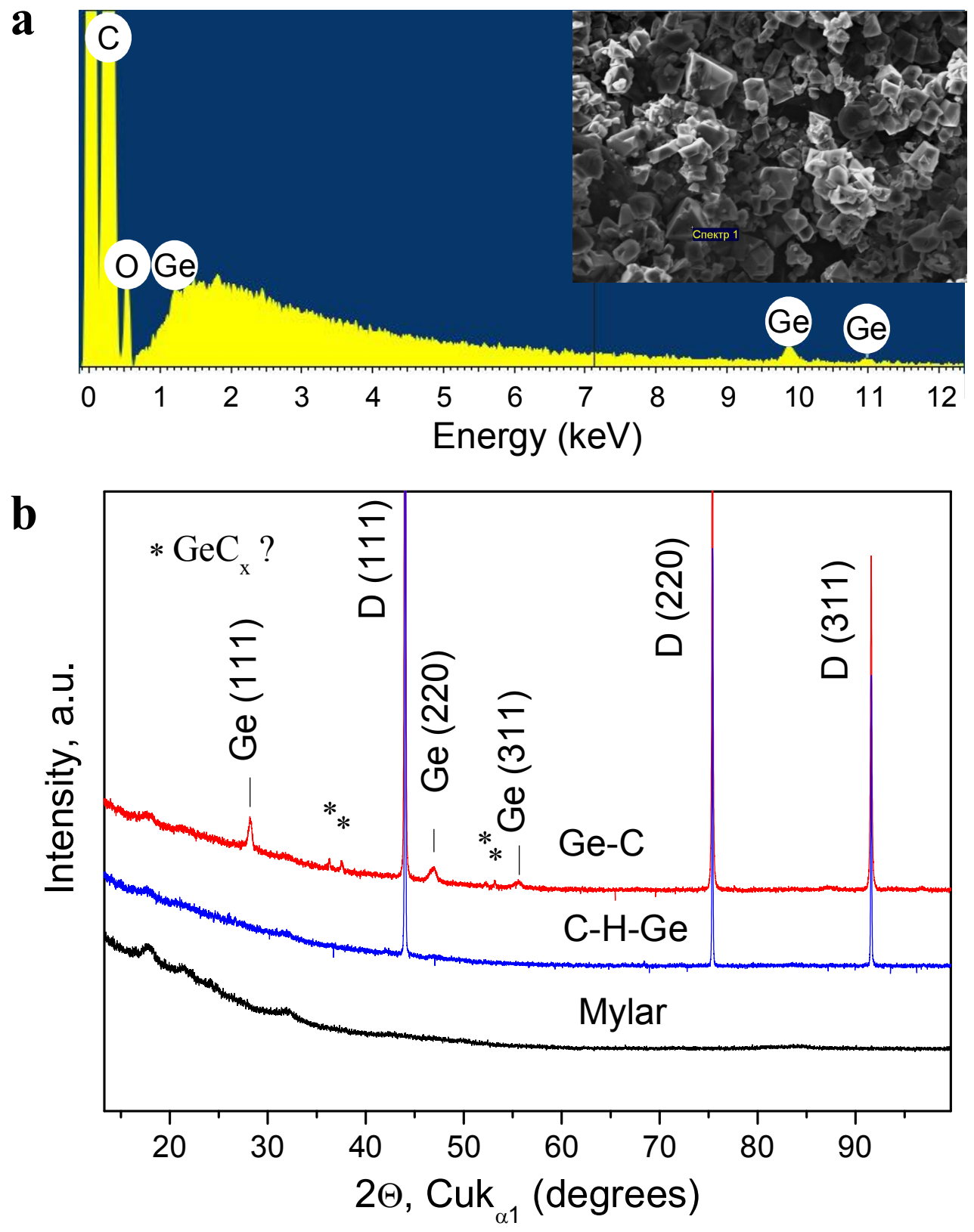

**Figure S1 | Ge-doped diamonds synthesized at high pressures. a**, Chemical composition of a diamond crystal, as-synthesized in the C-H-Ge growth medium. No metal-base impurities were detected by EDX microanalysis; silicon concentration in the reaction volume was below detecting limit, 0.1-0.2 %. **b**, X-ray diffraction analysis of acid-treated diamonds synthesized in the Ge-C and C-H-Ge growth systems. No secondary phases were found in diamond synthesized in the C-H-Ge system, while Ge-based inclusions were revealed in diamond produced in the Ge-C system. The diffraction lines of Ge (Fd3m, PDF#040545) are shifted to higher diffraction angles indicating the compressed state of germanium inclusions. Note that molar volume of germanium expands upon crystallization.

## 2. Optical measurements

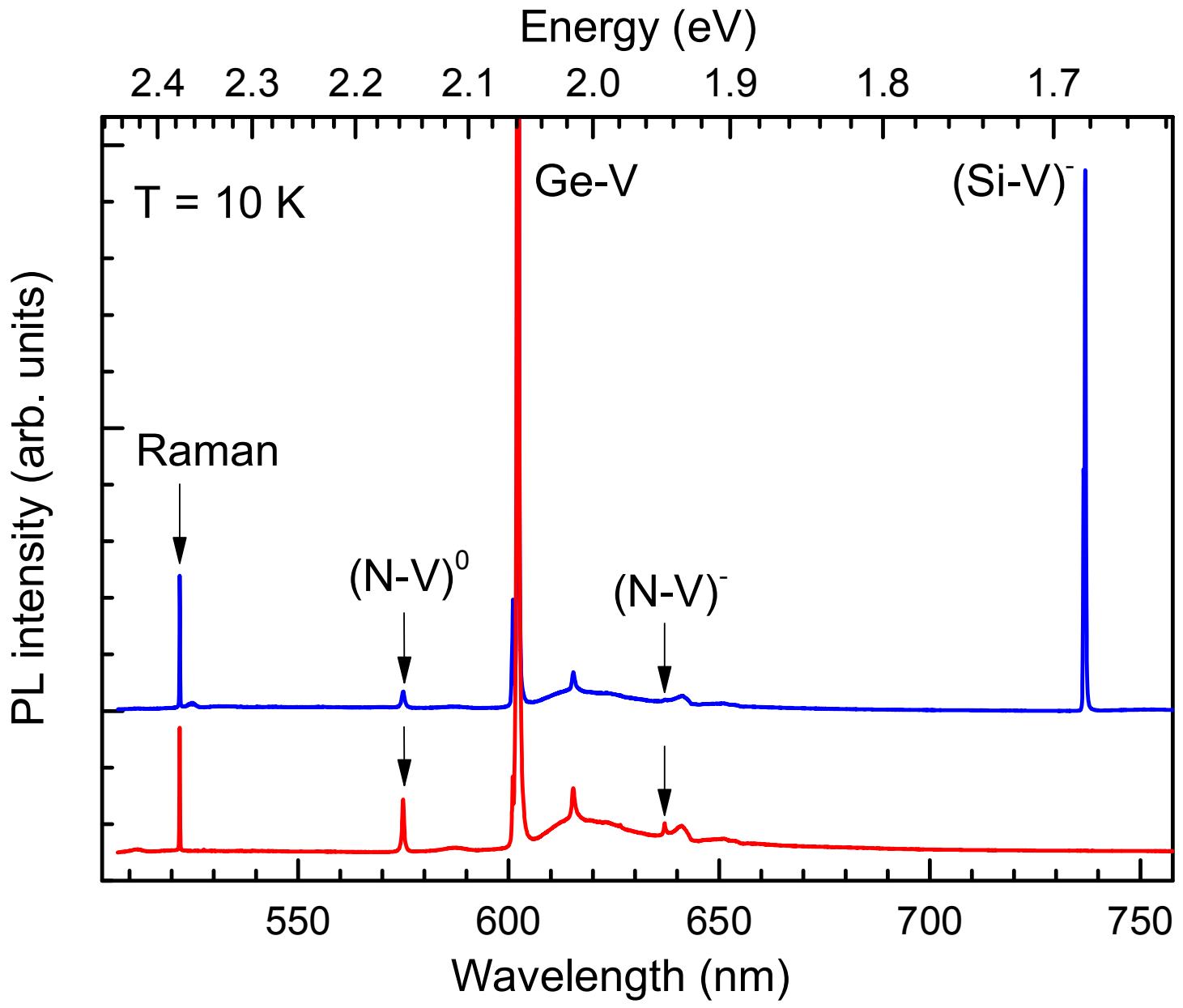

**Figure S2 | PL spectra at 10K with all main features**: Ge-related ZPL at 602 nm, the Raman peak from diamond, the $(N\text{-}V)^0$ and the $(N\text{-}V)^-$ peaks at 575nm and 637 nm, correspondingly, and the $Si\text{-}V^-$ peak at 737 nm. The presence of silicon impurity in our Ge-doped diamonds was easily eliminated by using a plexiglass mortar (the lower red spectrum) instead of a jasper one (the upper blue spectrum). The spectra are normalized to the intensity of the diamond Raman peak and displaced vertically for clarity.

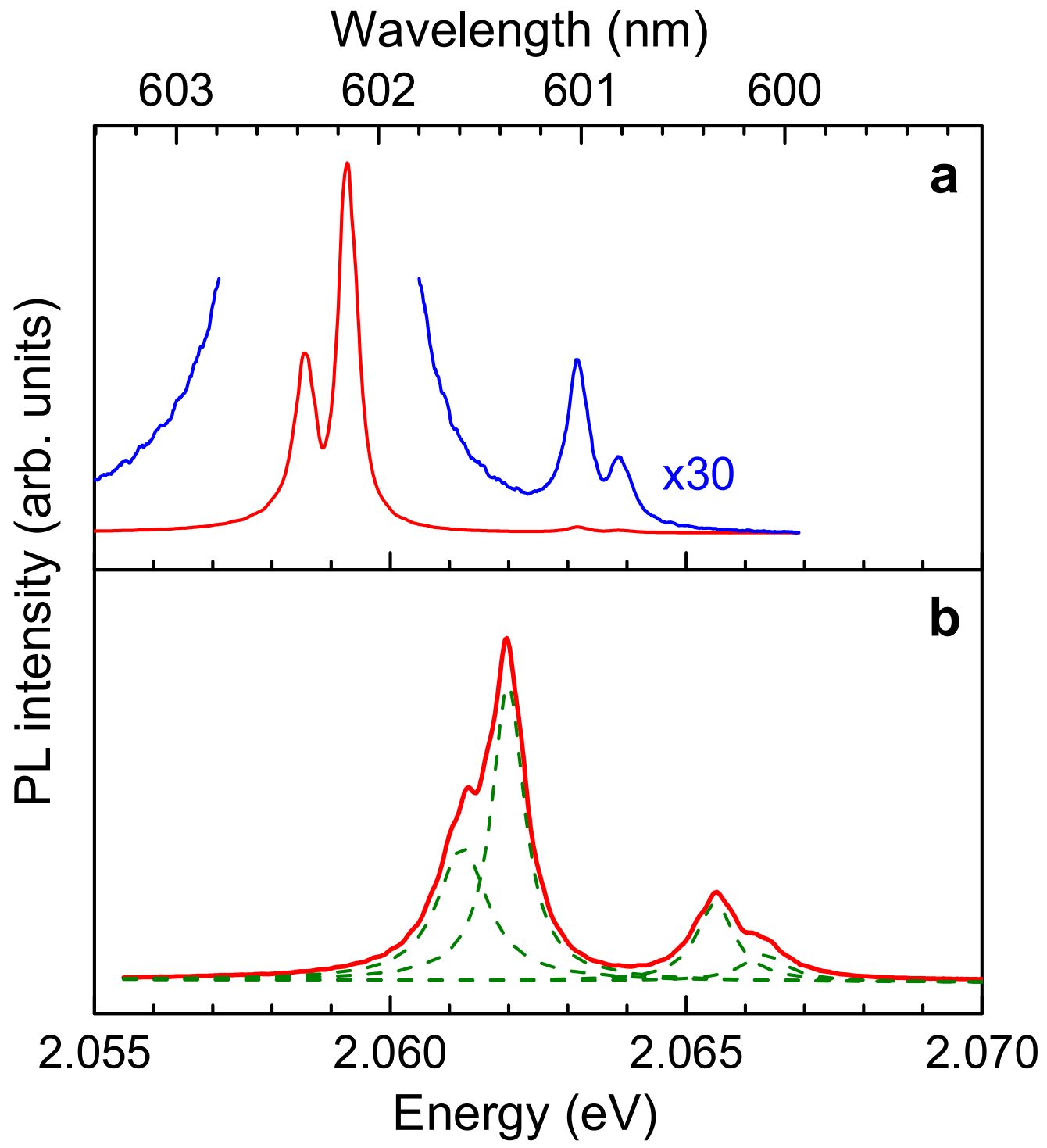

**Figure S3 | Carbon isotope shift in the fine structure of the ZPL**. On changing the $^{13}C$ content from 1.1% in natural carbon (**a**) to 99% (**b**), we observed isotope shift +2.8 meV for ZPL of the Ge-related colour centre. For the synthesis of diamond enriched with isotope $^{13}C$, about 12 mass % of $H_2O$ was added to the mixture of amorphous carbon powder with natural germanium.

### 3. About the effective masses.

The LVM peak for the samples with several isotopes of Ge is the sum of the contours from each isotope, with the appropriate weight. The observed isotopic shifts of ZPL and LVM for these samples are compared with the calculated ones in the framework of a local oscillator model, taking into account the effective masses $m*$ of the used germanium isotopes, introduced as follows:

$$\sqrt{\frac{1}{m^*}} = \sum_i n_i \sqrt{\frac{1}{m_i}}.$$

Here, $i$ runs over all isotopes of Ge, $n_i$ is the concentration of the $i^{th}$ isotope, and $m_i$ is the mass of this isotope[S1]. In the experiment, we used pure $^{72}$Ge (99.98%), Ge with natural abundance of the isotopes (36.5% $^{74}$Ge, 27.4% $^{72}$Ge, 20.5% $^{70}$Ge, 7.8% $^{73}$G, 7.8% $^{76}$Ge), and almost pure $^{76}$Ge (88% $^{76}$Ge, 12% $^{74}$Ge). Thus, the effective masses are $m_{72} = 71.9221$; $m^*_{nGe} = 72.6035$; and $m^*_{76} = 75.6407$, respectively.

### 4. Concentration of the (Si-V)$^0$ and (Si-V)$^-$centres, and estimates for Ge-V

In addition to the PL experiments, we have undertaken an attempt to register the absorption spectra of the Ge-V centres (see Figure S4). In order to increase the optical path, a KBr tablet was prepared with effective thickness of the sample 200 μm. As a result, when jasper mortar was used for the sample preparation, we were able to detect the absorption lines belonging to (Si-V)$^-$ and (Si-V)$^0$ with concentrations 7*10$^{15}$ cm$^{-3}$ and 8*10$^{15}$ cm$^{-3}$, respectively (according to [S2]). We observe the fine structure of the (Si-V)$^-$ centre at low temperatures, which also demonstrates a high quality of the obtained crystals. However, Ge-V lines in any of the measured samples were not observed. We assume that the absence of the absorption lines of the Ge-V centres can be related with a very low concentration of these centres. In this case, a strong luminescence of the Ge-V centre may indicate a high quantum yield of the luminescence. By a comparison with the luminescence and absorption of the (Si-V)$^-$ centres present in the same samples, we can give an upper estimate **< 5*10$^{14}$** cm$^{-3}$ for the concentration of the Ge-V centres.

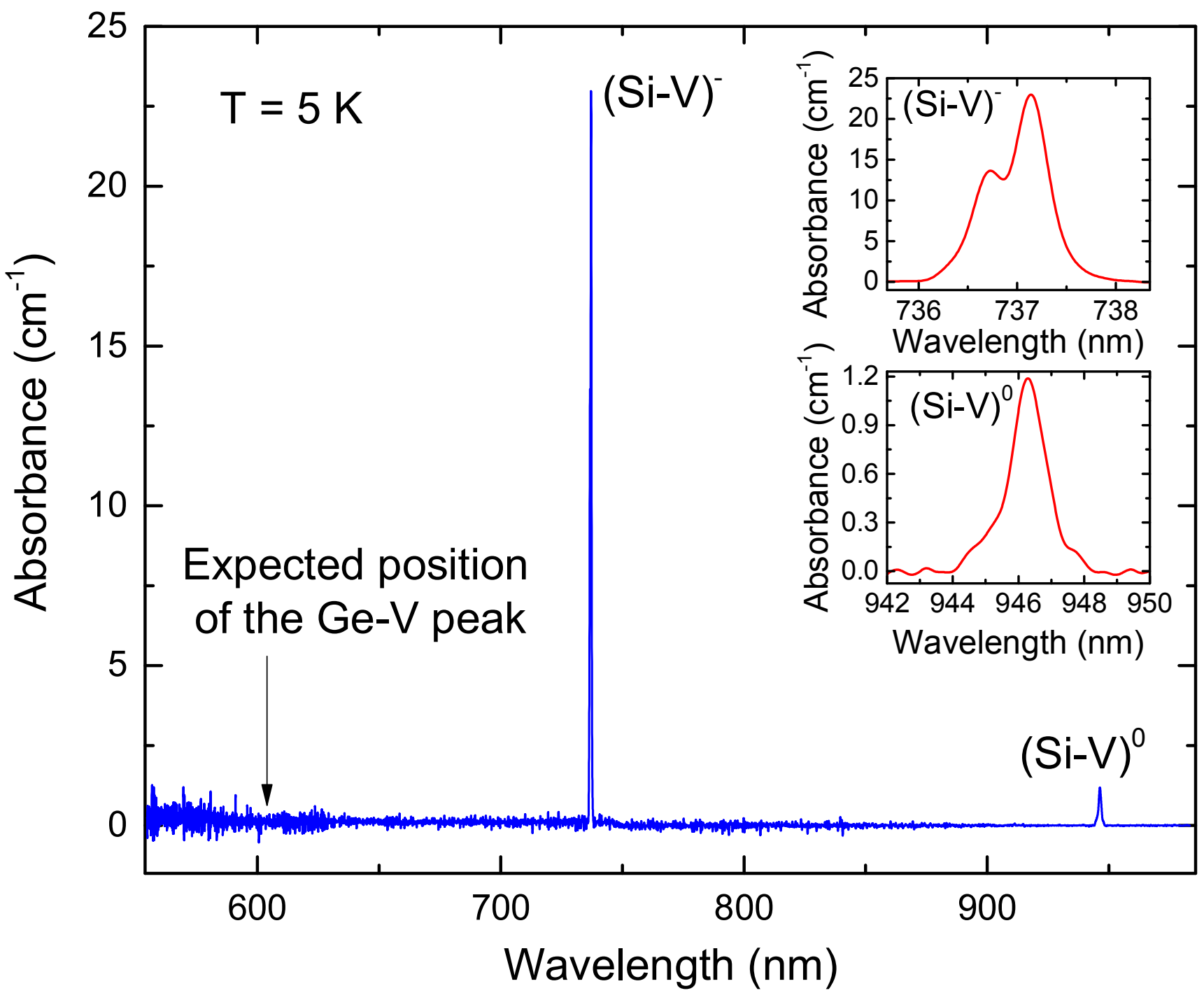

**Figure S4 | Absorption spectra of a diamond sample with Si impurity.** The peak absorption is 23 $cm^{-1}$ for the $(Si\text{-}V)^-$ centre and 1.2 $cm^{-1}$ for the $(Si\text{-}V)^0$ one. We didn't observe any absorption line near 602 nm, where the Ge-V peak could be expected.

## 5. Calculations

Basically, split-vacancy colour centre can be considered as two adjacent vacant sites (divacancy) with impurity atom introduced as substitutional defect which occupies one of these vacant sites. It follows from symmetry considerations that any divacancy in the diamond lattice should involve a lowering of the crystal symmetry from $Fd\bar{3}m$ to, at least, $R\bar{3}$. If one of the divacancy sites is occupied by a dopant (this geometry was taken as an initial step in the DFT calculation) the symmetry should be lowered even more (at least to$P3$). Taking into account these considerations, it seems natural to choose a supercell compatible with this broken symmetry. This supercell has some similarity to the crystal structure discussed before in [S3]. We should note the difference of our supercell geometry and ones used in previous works where the supercell was formed as two or three times multiple of original cubic cell in three directions. In the latter case the periodicity of supercell induces additional mirror symmetry of defects ordering.

The point group symmetry is$C_{3i}(\bar{3})$. Its irreducible representations are one-dimensional and degeneracy of defect levels are caused by the fact that $^1E_g/^2E_g$ on the one hand and $^1E_u/^2E_u$ on the other (corresponding to upper and lower pairs of defect levels, respectively) are complex conjugate, so they should have the same Kohn-Sham eigenvalues. As it was mentioned in the text, this "quasi"-degeneracy is lifted by taking into account either the spin-orbital or the Jahn-Teller effects.

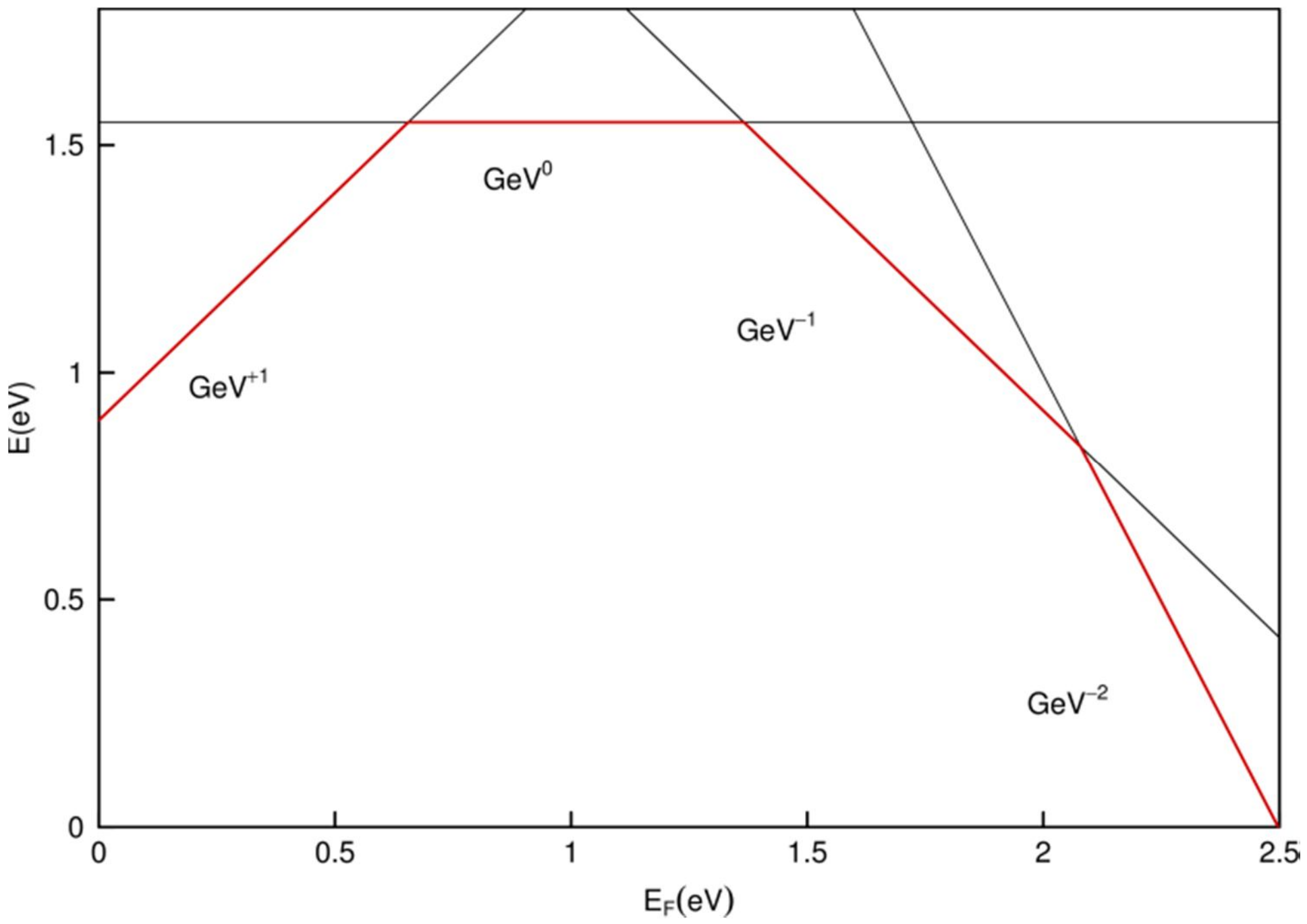

**Figure S5 | The calculated formation energy of the Ge-V defect as a function of the Fermi level position in the gap**. The crossing points represent the charge transition levels. The chemical potential of Ge is taken from the cubic germanium.

Calculation of defect formation energies demonstrates that four charged states (+1...-2) are possible (see Figure S5). It was found that $(Ge\text{-}V)^{-1}$ and $(Ge\text{-}V)^{0}$ are likely to exist in slightly p-doped diamond (which is quite realistic assumption, see discussion in the case of Si-V defects[S4]). It is interesting that changing of the charge state leads to a qualitative difference in the localization of the lower ($E_u$) defect levels which in a negatively charged impurity are located slightly above the valence band maximum. In this case, the excitation of electrons from these levels rather than from the valence band maximum should lead to the observed photoluminescence and absorption effects.

## 6. Supplementary References